\documentclass[a4paper,11pt]{article}
\pdfoutput=1 

\usepackage{jinstpub} 

\title{\boldmath Estimating the detection of antineutrinos in the future Brazilian neutron source}

\author[a,1]{Luiz Paulo de Oliveira,\note{Corresponding author.}}

\affiliation[a]{Instituto de Pesquisas Energ\'{e}ticas e Nucleares (IPEN/CNEN),\\ Av. Prof. Lineu Prestes, 2242 – Cidade Universit\'{a}ria – CEP 05508-000, S\~{a}o Paulo - Brazil}

\emailAdd{oliveira.phys@gmail.com}

\abstract{In line with the tradition of employing nuclear reactors for neutrino physics research, we present potential experiments that could be undertaken using the new Brazilian neutron source, the Brazilian Multipurpose Reactor (RMB). This upcoming facility has clearly defined objectives, including the production of radioisotopes for the healthcare system, material irradiation testing, and research using neutron beams to investigate the micro and mesostructures of various materials. However, we anticipate that the RMB will generate up to $3,500$ antineutrinos daily, potentially expanding the facility's scope to include neutrino physics experiments involving beta and inverse beta reactions. The modern operational conditions of the RMB and its location in the Iperó-Brazil region may also enable neutrino oscillation experiments and the determination of the $\theta_{13}$ parameter, as well as support safeguards activities by the International Atomic Energy Agency (IAEA), allowing for the monitoring of nuclear fuel burning cycles and isotope fractions. We discuss the potential utilization of the RMB as an antineutrino source, proposing experiments and addressing associated technical limitations.}


\begin{document}
\maketitle
\flushbottom

\section{Introduction}
\label{sec:intro}

Understanding the origin, destiny, and composition of the universe hinges upon unraveling the properties of neutrinos. Initially theorized by W. Pauli \cite{pauli} and E. Fermi \cite{fermi1}, neutrinos eluded detection until the late 1950s when Cowan and Reines successfully detected them in nature \cite{cowan}. This milestone was achieved through an experimental setup at the Savannah River nuclear reactor in South Carolina, where the phenomenon of inverse beta decay unveiled the existence of these elusive particles. Subsequently, neutrino experiments have undergone significant evolution, leading to discoveries regarding their properties. In the standard model of elementary particles, neutrinos are categorized as non-massive leptons, further divided into three flavors: electron, muon, and tau, interacting exclusively via the weak interaction. The elucidation of $W^{\pm}$ and $Z$ bosons played a pivotal role in comprehending beta decay, which yields electron antineutrinos as byproducts (Figure 1). The formidable challenge of the neutrino detection, initially highlighted by E. Fermi \cite{fermi2} in the 1930s, remains a prominent obstacle in contemporary scientific endeavors. However, the advancement of modern detectors has propelled neutrino experiments forward, contributing to our understanding of particle mass \cite{vogel}

The attribution of a finite mass to neutrinos serves as a compelling indication of physics beyond the standard model of elementary particles \cite{nakamura}, which currently undergoes numerous theoretical extensions, including supersymmetry \cite{miyazawa}, novel models for dark matter \cite{arbey}, and superstrings \cite{polchinski}, among others \cite{elko, witten}. Neutrino oscillation, a phenomenon of significant interest within the scientific community \cite{mcdonald, kajida}, led to the independent discoveries by physicists Arthur McDonald and Takaaki Kajita, meriting them the 2015 Nobel Prize in Physics. Despite this progress, several fundamental properties of neutrinos remain unanswered, including the mass hierarchy among the three flavors and the $\theta_{13}$ parameter, crucial for describing flavor oscillations \cite{kajida, vogel}. Presently, numerous large-scale international collaborations focus on neutrino experiments, such as NOvA \cite{nova}, MINOS \cite{minos}, and DUNE \cite{dune} at Fermilab-USA, T2K \cite{t2k} and Super-Kamiokande \cite{super} in Japan, Double Chooz \cite{double} and RICOCHET \cite{ricochet} in France, SNO \cite{sno} in Canada, and IceCube in South Pole \cite{ice}. In Brazil, two notable neutrino experiments are conducted in the city of Angra dos Reis, Rio de Janeiro state, namely the CONNIE collaboration (Coherent Neutrino-Nucleus Interaction Experiment) \cite{connie} and ANGRA \cite{angra}.

The CONNIE and ANGRA experiments are the only endeavors that use power reactor (the $4$ $GW$ Angra2) as neutrino source in the global Southern Hemisphere. CONNIE is dedicated to investigating coherent neutrino-nucleus interactions to unveil physics beyond the standard model, employing CCD detectors (Charge-Coupled Devices), sophisticated instruments capable of manipulating charges in small pulses \cite{connie2}. Initially focused on studying neutrino oscillation, ANGRA employed traditional tank-based techniques with ultrafiltered water, surrounded by photomultipliers \cite{angra2, angra3, angra4}. However, due to budget constraints, the collaboration shifted its focus to reactor core monitoring, utilizing a domestically produced detector. Leveraging the linear correlation between a nuclear reactor's thermal power and neutrino production \cite{russo}, the experiment facilitates real-time monitoring of Angra2, thereby contributing to the safeguard mission of the International Atomic Energy Agency (IAEA) \cite{agencia}.

Brazil is on the verge of constructing its new neutron source, the Brazilian Multipurpose Reactor (RMB) \cite{perrotta}. Situated in the city of Iperó, within the interior of São Paulo State, this research reactor will cater to the entire Brazilian populace, facilitating the production of radioisotopes for both the Unified Health System (SUS) and the private sector, alongside conducting irradiation tests on materials and engaging in neutron beam studies across a diverse array of techniques and materials \cite{pinho1}. The fuel element powering the RMB is domestically developed through the scientific endeavors of the Institute of Energy and Nuclear Research (IPEN) in São Paulo city, where seasoned researchers have formulated methodologies for constructing and analyzing the device \cite{perrotta2}, with a thermal power output of $30$ $MW$. Functioning as the dedicated facility for neutron-based instruments, the National Neutron Laboratory (LNN) will utilize thermal and cold neutron beams as probes to investigate the mesoscopic structure of matter, offering auxiliary laboratories for sample preparation and analysis, accessible to both industry and the broader Brazilian scientific community \cite{pinho1}. The LNN is poised to host a suite of $15$ instruments, including Small-Angle Neutron Scattering (SANS), featuring High-Intensity and High-Resolution Diffractometers \cite{pinho2, pinho4}, in addition to a Neutron Imaging Center \cite{eu}, among other facilities \cite{pinho1}. Currently, the project has advanced to the stage of finalizing detailed engineering designs and initiating groundwork on the designated site.

Retracing the historical milestone of neutrino detection in nuclear reactors achieved by Cowan and Reines, and drawing inspiration from the endeavors of the CONNIE and ANGRA experiments conducted in the Angra2 reactor, we assess the potential of the future Brazilian neutron source (RMB) to contribute to studies on neutrino properties. Although this objective was not initially part of the RMB project, this study explores potential applications of the reactor in this domain. This investigation aims to provide context for recent advancements in neutrino experiments (next section) and explore the potential contributions of the RMB to this field of research (third section), all while aligning with the primary objectives of the project. We will discuss possibilities and address physical and technological constraints in the concluding section.
\section{Neutrino Physics}

\subsection{Theoretical Aspects}

Neutrinos are elementary particles with zero electrical charge, leptonic quantum number $+1$ and that interact weakly like usual material. The phenomenon of neutrino oscillation, the exchange of flavors throughout propagation, arises from the fact that the eigenstates of flavor and mass are not the same. Similarly to the quark sector, where the Cabibbo-Kobayashi-Maskawa (CKM) matrix \cite{cabibbo, km} combines the basis of the states of down, strange and bottom quarks, in the neutrino sector there is a similar $3\times3$ matrix, called the Maki-Nakagawa-Sakata matrix (MNS) \cite{mns} relating the observable states $\nu_{k}$ where $k = e$, $\mu$ and $\tau$, with $\nu_{l}$ where $l = 1$, $2$ and $3$,
\begin{equation}
\begin{pmatrix}
\nu_{e}\\ \nu_{\mu}
\\ 
\nu_{\tau}
\end{pmatrix}=\begin{pmatrix}
U_{e1} & U_{e2} & U_{e3} \\ 
U_{\mu1} & U_{\mu2} & U_{\mu3} \\ 
U_{\tau1} & U_{\tau2}  & U_{\tau3}
\end{pmatrix}\begin{pmatrix}
\nu_{1}\\ \nu_{2}
\\ \nu_{3}

\end{pmatrix}.
\end{equation}
Currently, all elements of the MNS matrix were determined with good precision and approach unity, except for the element $U_{e3}$, which approaches $0$ and does not good precision. After the Chau-Keung parameterization \cite{chau}, using the convections $c_{ij} = cos(\theta_{ij})$ and $s_{ij} = sin(\theta_{ij})$, where $\theta_{ij}$ refers to the mixing angles between the neutrino mass states ($i,j = 1, 2, 3$), we can write
\begin{equation}
U_{MNS} = \begin{pmatrix}
1 & 0 & 0\\ 
0 & c_{23} & s_{23} \\ 
0 & -s_{23} & c_{23} 
\end{pmatrix} \begin{pmatrix}
c_{13} & 0 & s_{13}e^{i\delta} \\ 
0 & 1 & 0\\ 
-s_{13}e^{-i\delta} & 0 & c_{13}
\end{pmatrix} \begin{pmatrix}
c_{12} & s_{12} & 0\\ 
-s_{12} & c_{12} & 0\\ 
0 & 0 & 1
\end{pmatrix}
\end{equation}

\begin{equation}
\ = \begin{pmatrix}
c_{13}c_{12} & c_{13}s_{12} & s_{13}e^{i\delta}\\ 
-s_{12}c_{23}-c_{12}s_{23}s_{13}e^{-i\delta} & c_{12}c_{23}-s_{12}s_{23}s_{13}e^{-i\delta} & s_{23}c_{13}\\ 
s_{12}s_{23}-c_{12}c_{23}s_{13}e^{-i\delta} & -c_{12}s_{23}-s_{12}c_{23}s_{13}e^{-i\delta} & c_{23}c_{13}
\end{pmatrix}.
\end{equation}
We can observe that if the element $U_{13} << 1$, the value of $\theta_{13}$ is also small. The phase $\delta$ is responsible for the Charge-Parity (CP-symmetry) violation in the lepton sector, so any non-zero value of $\theta_{13}$ is required for a break.

Recently, the T2K experiment reported, for the first time, some indications of CP-symmetry violation in leptons \cite{t2k2}. Beams of muon neutrinos and antineutrinos were produced alternately by an accelerator. At the detector, a significantly greater proportion of electron neutrinos were detected from the muon neutrino beam compared to electron antineutrinos from the muon antineutrino beam. However, the results lacked the precision necessary to determine the extent of CP violation. Conversely, an experiment conducted by the NOvA collaboration found no evidence of CP violation in neutrino oscillations \cite{nova2}, indicating the need for further studies to confirm the phenomenon. 

Expressing the propagation of neutrinos as stationary states, that is, $\nu = \nu(0)e^{-iEt}$, we can obtain the transition probabilities \cite{reyna} between flavors parameterized in the Chaun-Keung scheme as
\begin{equation}
P(\nu_{e}\rightarrow \nu_{\mu})= P(\nu_{\mu}\rightarrow \nu_{e}) = sin^2(2\theta_{13})sin^2(2\theta_{23})sin^2\left (1.27\Delta m_{atm}^{2}\frac{L}{E}  \right ),
\end{equation}

\begin{equation}
P(\nu_{e}\rightarrow \nu_{\tau})= P(\nu_{\tau}\rightarrow \nu_{e}) = sin^2(2\theta_{13})cos^2(2\theta_{23})sin^2\left (1.27\Delta m_{atm}^{2}\frac{L}{E}  \right ),
\end{equation}

\begin{equation}
P(\nu_{\mu}\rightarrow \nu_{\tau})= P(\nu_{\tau}\rightarrow \nu_{\mu}) = sin^2(2\theta_{23})cos^4(2\theta_{13})sin^2\left (1.27\Delta m_{atm}^{2}\frac{L}{E}  \right ),
\end{equation}
where $E$ and $L$ are energy and propagation distance of neutrinos and, $\Delta m_{ij}^{2}$ is the difference of squared masses of two states $i,j$. Good Super-Kamiokande data suggests a large value for $\Delta m_{atm}^{2} = 2 \times 10^{-3} eV^2$, and we therefore use this value as an upper approximation ($\Delta m_{ij}^{2} \approx  \Delta m_{atm}^{2}$). From equations ($2.4 - 2.6$) we can observe the dependence on $\theta_{13}$, so that the oscillation of neutrinos between the flavors and the symmetry of the exchange $P(\nu_{\mu}\rightarrow \nu_{\tau})$ for $P(\nu_{\tau}\rightarrow \nu_{\mu})$ occurs. We highlight that the oscillation ceases to exist in the limit $E >> 1.27m_{atm}^{2}L$, limited to effect only for low-energy neutrinos.

\begin{figure}
    \centering
    \includegraphics[scale = 0.5]{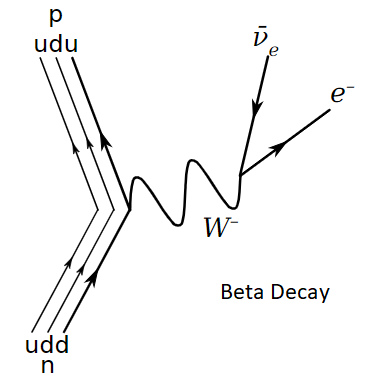}
    \caption{Beta decay at the quark level is the mediator of the weak interaction $W^{-}$.}
\end{figure}

\subsection{Detectors}
Due to their rare interactions with matter, neutrino detectors must be either significantly large or exceedingly sensitive to detect substantial numbers of events. One common detection method involves employing large tanks filled with purified water and equipped with photomultipliers to detect Cherenkov radiation (appearing as blue light), which results from neutrino interactions with water, producing muons. Complemented by vertical and horizontal tunnels, these tanks utilize natural elevation shielding to filter out cosmic neutrinos, minimizing interference in the recorded results. An exemplary implementation of this detection approach is the Super-Kamiokande experiment, situated at a depth of $1$ km. It consists of a cylindrical tank containing $50,000$ tons of ultrapure water to facilitate particle stopping, with the trajectory of particles measured by $11,129$ internal photomultiplier tubes \cite{super}.

Another detection method involves utilizing a liquid scintillator (LS) doped with gadolinium. For instance, the Double-Chooz experiment \cite{double} utilizes a detector comprising $10$ $m^{3}$ of liquid scintillator specifically designed for this purpose. The liquid scintillator is doped with gadolinium to tag neutrons resulting from inverse beta decay triggered by antineutrinos emitted by the nuclear reactor. Surrounding the detector's target are layers of other liquids that shield it from external particles and environmental radioactivity. Within the liquid scintillator, $390$ photomultiplier tubes are submerged, converting interactions into electronic signals.

Solid-state detectors have made significant advancements owing to new material technologies and increased production capacity. These detectors can capture electrical signals emitted by charged particles in interactions involving neutrinos, allowing for the precise determination of the energies and trajectories of these elusive particles. The CONNIE Collaboration exemplifies an experiment utilizing solid-state detectors to investigate neutrino-related reactions \cite{connie}. CONNIE employs state-of-the-art skipper-CCD technology, developed in laboratories at LBL and Fermilab (USA), which significantly reduces detector noise. This technology enables the individual detection of electrons, rendering it a quantum detector with unparalleled sensitivity.

The neutrino detection scheme thus relies on the type of source (cosmic, nuclear reactors, and accelerators) as well as the specific characteristics of each facility. However, the sensitivity of radiation and charged particle detectors is a \textit{sine qua non} condition for ensuring the quality of data from experiments of this nature in the forthcoming decades.

\section{Future Brazilian Neutron Source}

The RMB will constitute a new division of the National Nuclear Energy Commission (CNEN) and will mark the onset of a new phase in nuclear research in Brazil. Comprising a core assembly of $5\times 5$ elements of $U_{3}Si_{2}$ (19.75 wt percent) fuel, each element consists of $21$ parallel plates encased in Aluminum $6061$ alloy. The neutron flux generated within the reactor core approximates $1\times 10^{14}$ $n/cm^{2}s$, as per the Monte Carlo simulation depicted in Figure 2. The reflector tank is strategically designed to facilitate irradiation tests on materials, serve as a source of cold neutrons, and accommodate five neutron collector tubes—two designated for transporting cold neutrons and three for thermal neutrons. Initially, during the initial construction phase, the suite of $15$ instruments will be powered by two thermal and one cold neutron beam. For instance, the neutron imaging center enables the examination of material structures utilizing tomography techniques from positions merely $5$ meters away from the reactor core \cite{eu}. In essence, the planned configuration permits studies with assemblies positioned between units and tens of meters away from the reactor core.  

\begin{figure}
    \centering
    \includegraphics[scale = 0.25]{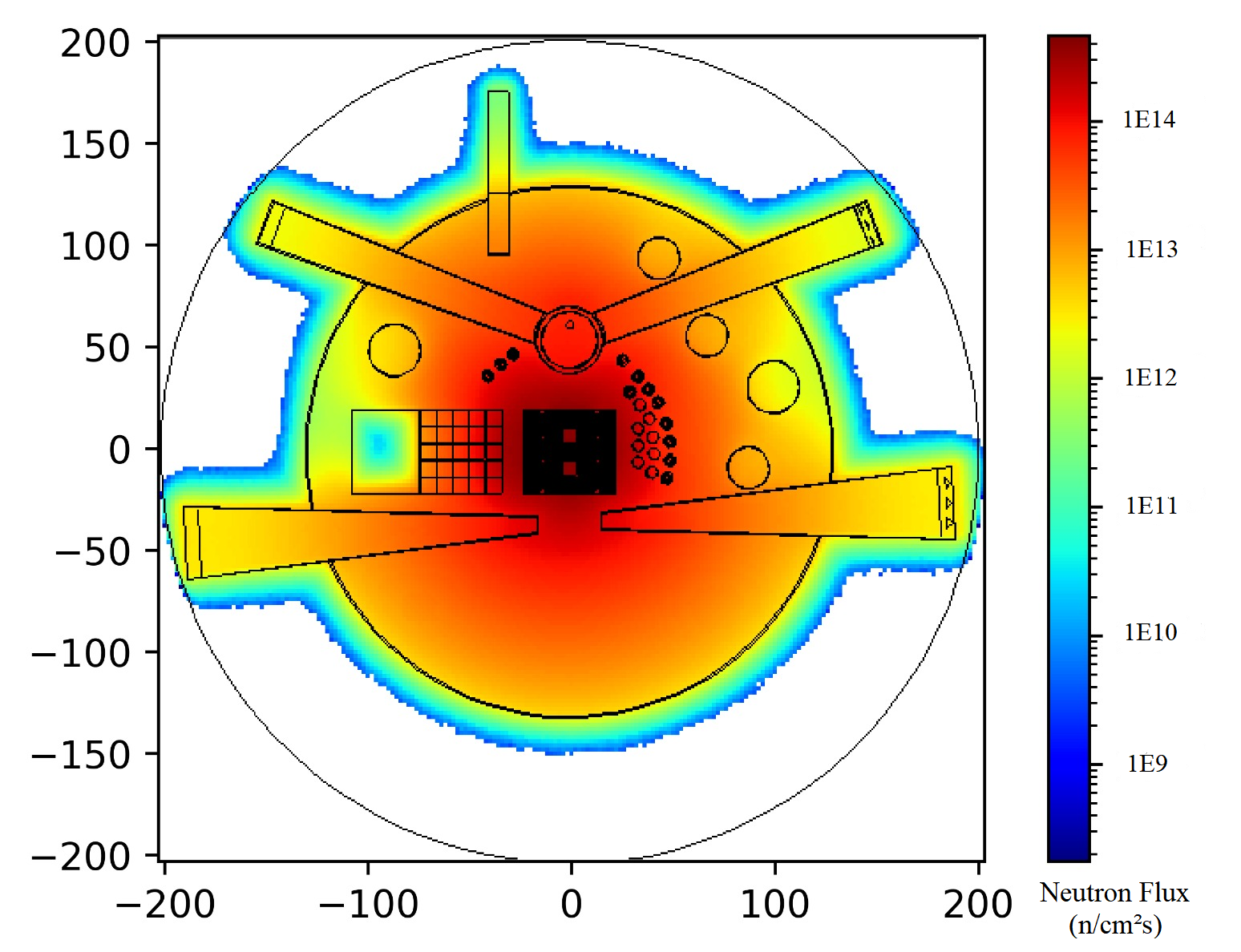}
    \caption{Monte Carlo simulation of the RMB reflector tank with dimensions X-Y in $cm\times cm$.}
\end{figure}

The RMB complex is situated adjacent to the Brazilian Navy Experimental Center, which stands $600$ meters tall, as illustrated in Figure 3. In this same vicinity, Morro do Araçoiaba, a natural elevation, can be observed, reaching a maximum height of $980$ $m$ above sea level. The region surrounding Morro do Araçoiaba comprises arenite at its highest point and $Fe_{2}O_{3}$ at its lowest, where iron exploration deposits are located.

\begin{figure}
    \centering
    \includegraphics[scale = 0.7]{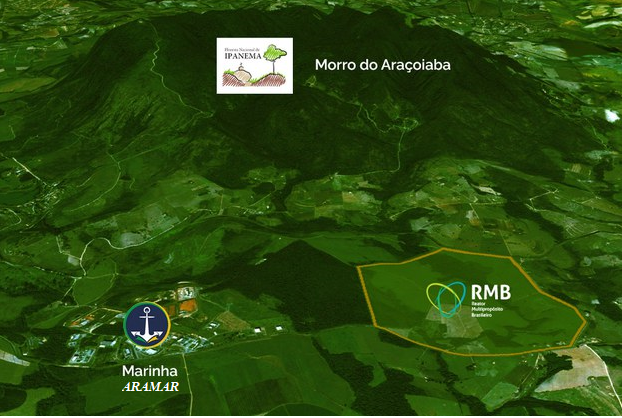}
    \caption{Region of the future RMB, next to the ARAMAR Experimental Center of the Brazilian Navy and Morro do Araçoiaba in the background \cite{site}.}
\end{figure}
A good approximation for calculating the rate of neutrino interactions in water is given by $R_{\nu} = N_{f}N_{p}\left \langle \sigma  \right \rangle/4\pi D^{2}$, where $N_{f}$ is the average fission rate given by
\begin{equation}
N_{f}(events/s)=6.24\cdot 10^{-8}P_{therm}(MW)/E_{fission}(MeV),
\end{equation}
where $P_{therm}(MW)$ is the thermal power of reactor, $E_{fission}(MeV)$ is the energy per fission, $N_{p}$ is the number of protons in the fiducial volume, $\left \langle \sigma  \right \rangle$ is the average cross section, $D$ is the distance between source and detector \cite{angra4}. In the case of RMB we have $R_{\nu}(events/day) \approx  2.411\times 10^{4} \times V(m^{3})/ (D(m))^2$, where $V(m^{3})$ is the fiducial volume. LNN has a large structure to house neutrino detectors within its hall ($D < 50$ m). If necessary, accommodation outside the LNN building could be built to house the experiment. However, the possibility of implementing a detector of volume $V \approx 3.4 $ $m^{3}$ and $D \approx 5$ $m$ is viable in the vicinity of the neutron imaging instrument \cite{eu}. This configuration would offer $R_{\nu} = 3279$ events/ day.

While neutrino detection is not among the primary objectives and purposes of the RMB, it is undeniable that a neutron source boasting excellent personnel and operational infrastructure can potentially inspire other applications without detracting from its intended activities. In this vein, this study aims to assess potential neutrino detection experiments leveraging the infrastructure envisioned in the RMB project. Drawing inspiration from the concepts explored in the Brazilian ANGRA experiment and incorporating advancements in particle detection technology, we discuss potential applications in the following section.

\section{Experiment possibilities}

\subsection{Neutrinos oscillation}
The probability of survival of electron antineutrinos in vacuum \cite{reyna}, to first order, is given by 
\begin{equation}
    P(\bar{\nu _{e}}\rightarrow \bar{\nu _{e}}) = 1 - sin^2(2\theta _{13})sin^2\left (1.27\Delta m _{atm}^{2}\frac{L}{E}  \right )-cos^4(\theta _{13})sin^2(2\theta _{12})sin^2\left (1.27\Delta m_{12}^{2}\frac{L}{E}  \right ),
\end{equation}
where we can observe that $sin^{2}(2\theta_{13})$ behaves as an amplitude of the second term after equality, while its contribution in the third term is almost non-existent. An adjusted graph with the best parameters ($sin^2(2\theta_{12})=0.8$, $\Delta m_{atm}^{2}=2\times10^{-3}$ $eV^{2}$, $\Delta m_{12}^{2}=8\times10^{-5}$ $eV^{2}$ ) can be found in Figure 4, where we use $sin^{2}(2\theta_{13})=0.10$, an excellent ansatz according to the most recent limits \cite{espectro}. The vicinity of the RMB may allow experiments on neutrino oscillations, given the presence of natural elevations that serve as natural shielding for cosmic rays. However, we must point out that the reactor is already $600$ m above sea level and therefore, we would only have a net effect of $980 - 600 = 300$ m of shielding. Aretine has an average density of $2.5$ $g/cm^{3}$, while iron oxide has $5.4$ $g/cm^{3}$, providing approximate maximum shielding values of $780$ m.w.e and $1620$ m.w.e, respectively.

Due to the effects of flavor oscillations, we classify the interval $0 \leq L \leq 4$ Km as the zone of interest. In fact, the inner regions of Morro do Araçoiaba have distances close to the range, but the higher altitude regions exceed the range $0 \leq L \leq 4$ Km, and therefore, would harm the detection of antineutrinos. The vicinity of the reactor does not produce natural shielding to minimize the effects of muons.
According to Mei and Hime \cite{muon}, the intensity of the muon beam arriving at Earth for a given altitude in meters of water equivalent (m.w.e) is given by
\begin{equation}
I_{\mu }(h_{0}) = 67.97\cdot 10^{-6}\cdot e^{-\frac{h_{0}}{0.285}}+2.071\cdot 10^{-6}\cdot e^{-\frac{h_{0}}{0.698}},
\end{equation}
where $I_{\mu}(h_{0})$ is given by $cm^{-2}s^{-1}$. Therefore, in the absence of natural shielding, there would be a need for shielding developed in many layers and the use of LC. The detection of neutrinos by the CONNIE collaboration using skipper-CCDs has been breaking records with data increasingly free of external noise, and the technique used could be applied to a flavor oscillation experiment in the future.

An expansion of the trigonometric functions for small arguments in equation $(4.1)$ gives the relation $P(\bar{\nu _{e}}\rightarrow \bar{\nu _{e}}) \approx 1 - C\left (\frac{L}{E}  \right )^2,$ where $C = 6.45(\theta _{13}^{2}\Delta _{atm}^{2}+\theta _{12}^{2}\Delta _{12}^{2}).$ From this exercise, we can observe that the probability of detecting antineutrinos depends quadratically on the parameter $\theta_{13}$ and the distance $L$. The larger $\theta_{13}$ and $L$, the lower the probability of antineutrino survival. Therefore, we conclude that implementing flavor oscillation experiments in the RMB has a small parameter window and great dependence on new neutrino detection technologies.

\begin{figure}
    \centering
    \includegraphics[scale = 0.5]{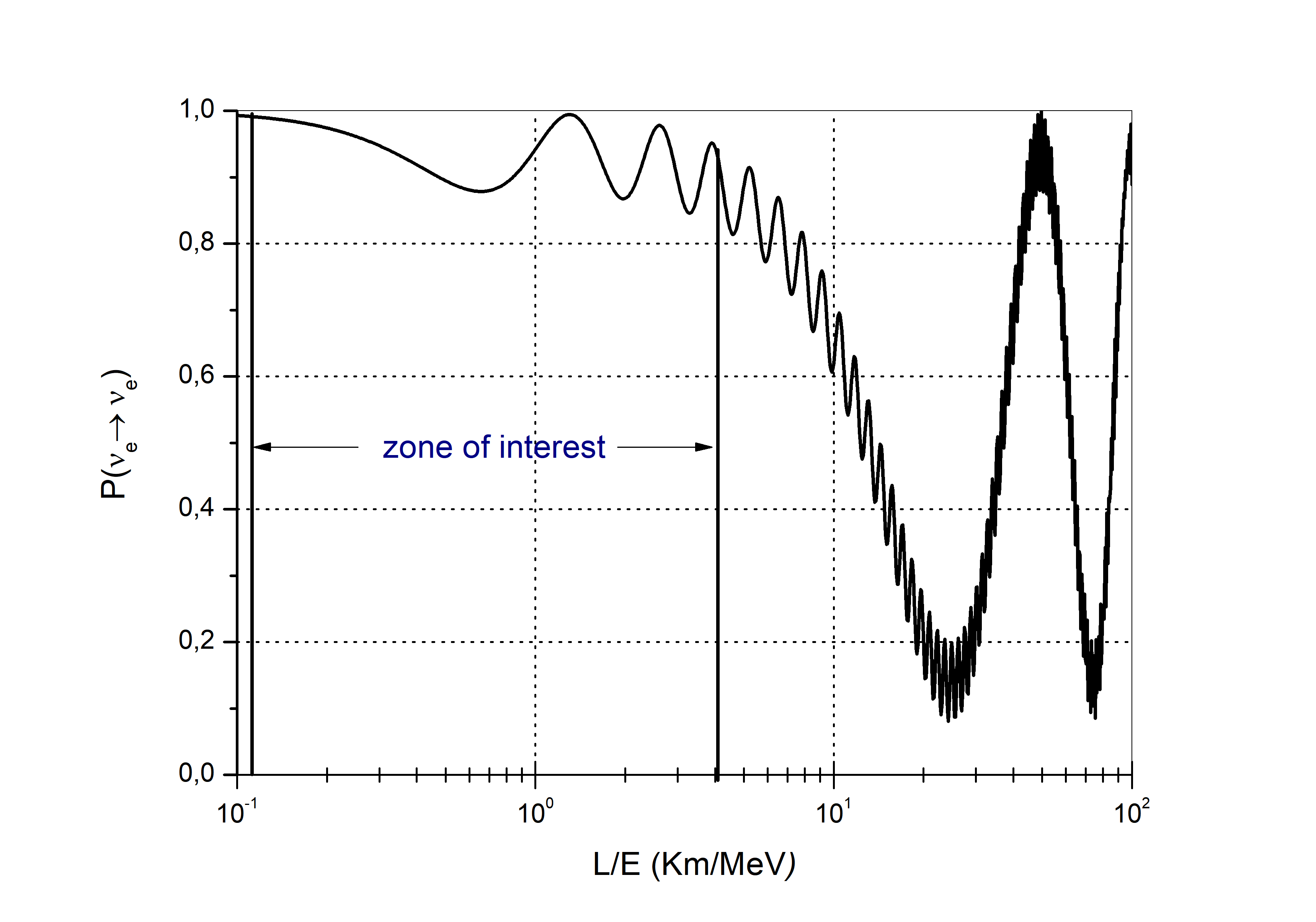}
    \caption{Electron antineutrino survival probability as a function of ratio $L/E$ (Km/MeV). The area of interest is highlighted $L < 4$ km.}
\end{figure}

\subsection{IAEA Safeguards}
Safeguards encompass activities through which the International Atomic Energy Agency (IAEA) can verify a State's compliance with its commitment to refrain from utilizing nuclear programs for military purposes, serving as a cornerstone within the global nuclear non-proliferation regime. The safeguards system operates as a confidence-building measure \cite{site2}. In this context, given that the number of events per unit of time is directly proportional to the reactor's power, it becomes feasible to monitor the consumption of core fuel through the flux of antineutrinos generated in beta decay. It is well-established that, on average, $6\nu_{e}$ per fission are emitted, penetrating the reactor containment and surrounding structures unhindered, with energies ranging up to $8$-$10$ MeV \cite{safe}.  

In addition to assessing the reactor's power output based on the number of detected events, some studies propose that the composition of isotopes within the fuel element can also be monitored using the proposed methodology. This is feasible because the energy spectrum of neutrinos varies for each isotope \cite{huber}, allowing for the monitoring of individual fractions of the nuclear fuel. Last year, the Double Chooz experiment obtained data on the electron antineutrino spectrum, which we present in Figure 5 \cite{espectro}. Hence, detecting electron antineutrinos can serve as an effective method for monitoring the fuel-burning process in the RMB.
\begin{figure}
    \centering
    \includegraphics[scale = 0.5]{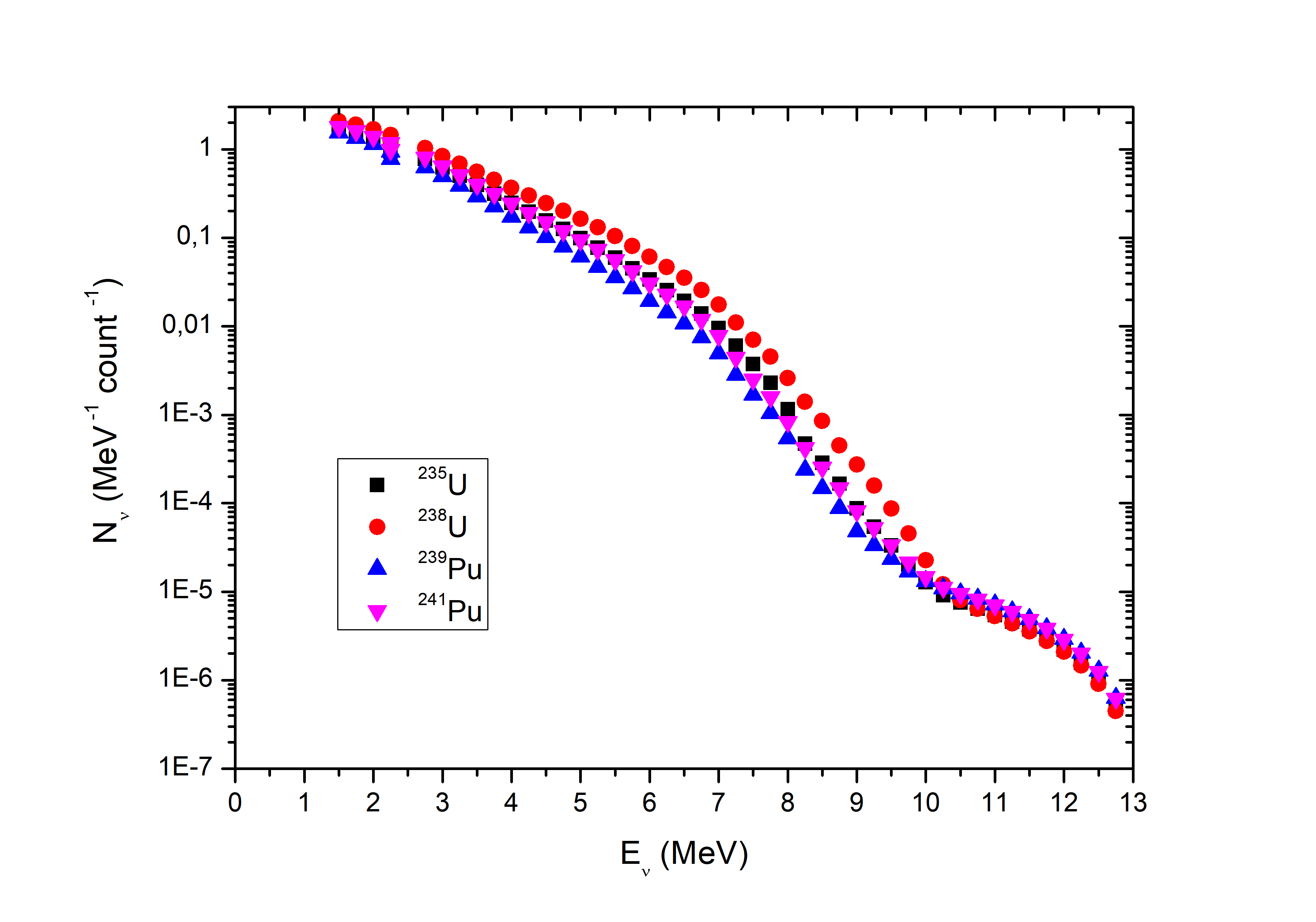}
    \caption{Spectra of antineutrino of $^{235}U$, $^{238}U$, $^{239}Pu$ and $^{241}Pu$ for two years of fuel irradiation taken from the Double Chooz Experiment \cite{espectro}.}
\end{figure}

\section{Conclusions}
We have outlined several potential experiments involving neutrinos that could be conducted with the construction of the upcoming Brazilian neutron source, the RMB. In addition to the objectives set forth by the new CNEN unit, the $30$ $MW$ reactor offers favorable operational conditions for detecting electron antineutrinos emitted from nuclear fission reactions.

We estimate that the RMB can detect approximately $3,500$ events per day, an amount competitive with other research reactors. The modern operating conditions of the research center enable investigations into neutron physics from distances of mere meters away from the RMB core. These conditions facilitate research into neutrino flavor oscillations, including the determination of the $\theta_{13}$ parameter, thereby addressing the issue of neutrino mass hierarchy. Moreover, the geographical location of the facility provides strategic opportunities for utilizing the natural shielding provided by Morro do Araçoiaba to mitigate the impact of cosmic muon flux. Although initially insufficient, this natural shielding can be reinforced through the development of novel shielding methods and by enhancing the capacity and sensitivity of new detectors.

Another significant application of the RMB's antineutrino source lies in its potential role in nuclear weapons non-proliferation safeguards overseen by the IAEA. Antineutrino detection can facilitate the monitoring of nuclear fuel burning within each isotopic fraction, thereby enhancing reactor safety and operational oversight. Similar to the Double Chooz experiment, the RMB will have the capability to position detectors close to the reactor core, enabling investigations into stereo neutrinos \cite{stereo}, thus shedding light on advancements in physics beyond the standard model of elementary particles.

\acknowledgments

This work is dedicated to the memory of Edson Lucas de Oliveira, a young engineer, and brother who fulfilled his mission and left a legacy of serenity, honor, and dignity. We will always love you.
 
\bibliographystyle{JHEP2}
\bibliography{main}










\end{document}